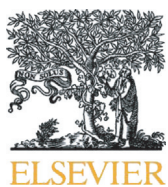
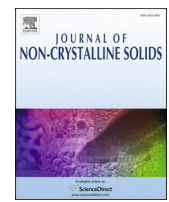

# Medium range structural order in amorphous tantala spatially resolved with changes to atomic structure by thermal annealing

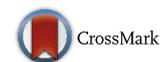

Martin J. Hart [a,*], Riccardo Bassiri [b], Konstantin B. Borisenko [c], Muriel Véron [d], Edgar F. Rauch [e], Iain W. Martin [a], Sheila Rowan [a], Martin M. Fejer [b], Ian MacLaren [a]

[a] SUPA, School of Physics and Astronomy, University of Glasgow, G12 8QQ, United Kingdom
[b] E. L. Ginzton Laboratory, Stanford University, Stanford, CA 94305, USA
[c] Department of Materials, University of Oxford, Parks Road, Oxford OX1 3PH, United Kingdom
[d] Univ. Grenoble Alpes, SIMAP, F-38000 Grenoble, France
[e] CNRS, SIMAP, F-38000 Grenoble, France

## ARTICLE INFO

## ABSTRACT

Amorphous tantala (a-Ta$_2$O$_5$) is an important technological material that has wide ranging applications in electronics, optics and the biomedical industry. It is used as the high refractive index layers in the multi-layer dielectric mirror coatings in the latest generation of gravitational wave interferometers, as well as other precision interferometers. One of the current limitations in sensitivity of gravitational wave detectors is Brownian thermal noise that arises from the tantala mirror coatings. Measurements have shown differences in mechanical loss of the mirror coatings, which is directly related to Brownian thermal noise, in response to thermal annealing. We utilise scanning electron diffraction to perform a modified version of Fluctuation Electron Microscopy (FEM) on Ion Beam Sputtered (IBS) amorphous tantala coatings, definitively showing an increase in the medium range order (MRO), as determined from the variance between the diffraction patterns in the scan, due to thermal annealing at increasing temperatures. Moreover, we employ Virtual Dark-Field Imaging (VDFi) to spatially resolve the FEM signal, enabling investigation of the persistence of the fragments responsible for the medium range order, as well as the extent of the ordering over nm length scales, and show ordered patches larger than 5 nm in the highest temperature annealed sample. These structural changes directly correlate with the observed changes in mechanical loss.

© 2016 The Authors. Published by Elsevier B.V. This is an open access article under the CC BY license (http://creativecommons.org/licenses/by/4.0/).

## 1. Introduction

Ion-beam sputtered amorphous tantala (a-Ta$_2$O$_5$) is often the material of choice for the high refractive index layer of highly reflective thin film coatings and find widespread applications that range from optical atomic clocks [1], ring laser gyroscopes [2], frequency comb techniques [3] and high-precision interferometers such as the Laser Interferometer Gravitational-wave Observatory (LIGO) [4]. Amorphous tantala also has applications that include insulating films with high dielectric constant for electronics [5] and corrosion resistant coatings for biomedical applications [6].

However, the performance of the coatings must be improved to make them a viable option for future upgrades to ultra-high precision gravitational wave interferometers, which are currently expected to be limited, at their most sensitive frequencies, by thermal noise arising from the coatings. To do so, it is necessary to understand changes in the atomic structure that occur during manufacturing and post-processing. Previous studies have shown that doping and annealing of the thin films cause considerable changes to the macroscopic properties such as optical absorption, scattering and mechanical loss [7–9]. Mechanical loss is equivalent to internal friction and is defined as the reciprocal of the mechanical Q factor, a quantity that describes the level of damping in an oscillator; a higher Q value indicates a lower rate of energy loss per oscillation with respect to the energy that is stored within the oscillator. This measure of mechanical stability and its relationship to Brownian thermal noise is quantified through the fluctuation dissipation theorem by Callen and Greene [10]. In general, only small changes are observed in the atomic structure of the coatings of the same material prepared by different methods when studied by electron diffraction reduced density function (RDF) analysis (which appear to be only sensitive to short range order, principally around the 1st and 2nd nearest neighbours, and up to a maximum of about 1 nm, and consequently cannot distinguish between atomistic models that contain or do not contain nanoscale order) [11–13]. A previous study using this technique has however demonstrated a correlation between mechanical loss and the concentrations of titania in titania-doped tantala [14]. Extended X-ray

* Corresponding author at: SUPA, School of Physics and Astronomy, University of Glasgow, Kelvin Building Rm 427a, G12 8QQ, United Kingdom. Tel.: +4401413301658.
E-mail address: m.hart.1@research.gla.ac.uk (M.J. Hart).

Absorption Fine Structure (EXAFS) shows more extensive order to perhaps the 3rd or 4th nearest neighbour and is chemically sensitive to the ordering about specific atoms, although this still does not reveal any order beyond 1 nm. Our recent work has shown that this provides further insights into the behaviour on annealing [15]. The previous electron diffraction studies have shown the atomic structure to be homogeneous at volumes probed with electron beams of 50 nm in lateral extent up to a 600 °C heat treatment, whereas in the present study the atomic structure is shown to be heterogeneous over volumes probed with an electron beam of 2 nm in lateral extent. It has been suggested previously, that this apparent homogeneity of the coating atomic structure is a consequence of the scale at which the structure has been examined; the inherent averaging over volumes containing hundreds of millions to hundreds of billions of atoms averages out any local structural differences in the materials. It is then expected for there to be a peak in the heterogeneity of the atomic structure at a scale determined by the volume of the material probed, coinciding with the maximum variance in structural order. In this work, the volume of material probed contains in the order of thousands of atoms, and proper variation in the ordering of the structure at the medium range can be quantified.

Fluctuation Electron Microscopy (FEM) is a diffraction and/or imaging technique that quantifies medium range order (MRO) in the roughly 1 to 3 nm range. The original formulation of FEM Gibson and Treacy [17] examines the MRO by measuring spatially resolved diffracted intensity fluctuations from nano-volumes in the sample material through the normalised variance,

$$V(k,Q) = \frac{\langle I^2(\mathbf{r},k,Q)\rangle}{\langle I(\mathbf{r},k,Q)\rangle^2} - 1, \tag{1}$$

where $I(\mathbf{r},k,Q)$ is the diffracted intensity as a function of position $\mathbf{r}$ on the sample, scattering vector $k$, probe size $Q$, and $\langle\ldots\rangle$ indicates averaging over $\mathbf{r}$. The technique is sensitive to three- and four-body correlations [16], and the fluctuations are maximally sensitive when the electron probe size is of comparable length scale to the MRO structural ordering being probed. So the extent of MRO is quantified through the magnitude of the variance of the diffracted intensity, as a function of scattering vector over a length scale determined by the size of probe used. Originally proposed by Gibson and Treacy [17], the technique was initially carried out using dark-field imaging in the TEM, although an equivalent experiment can be carried out using scanning diffraction [18]. This latter experimental approach has distinct advantages, especially on modern scanning transmission electron microscopes, where probes well below 1 nm in diameter can be routinely produced, provided the diffraction patterns can be acquired reasonably quickly (which is now possible due to advances in imaging detectors).

Atomistic models have shown that the variance displays clear trends as a function of the size and volume fraction of the ordered regions [19, 20] and, to date, the technique has been employed to show variation in the nanoscale order of amorphous silicon [21–26] and amorphous germanium [16, 18, 27] thin films, phase change chalcogenide materials [28–30], and a selection of amorphous metals [31, 32]. In these experiments, qualitative differences in FEM variance were observed and attributed to fundamental physical phenomena such as differences in film deposition condition [23], the existence and thermal ripening of subcritical nuclei that precede crystallisation [29, 30], and the effect of alloying on crystallisation kinetics [33]. Quantitative FEM analysis has thus far proven challenging, but with recent developments such as variable resolution FEM, information about the extent of the nanometre-scale ordering can be extracted [26, 34]. Nevertheless, a number of recent studies have been successful in relating the scattering covariance and angular correlations in FEM data to structural information [13, 35, 36].

In recent work, we used scanning nano-diffraction FEM to collect data, in a similar way to that described by Voyles and Muller [16], and demonstrated the existence of MRO in a-Ta$_2$O$_5$ [37]. In the version of FEM applied in the present work, we depart from the standard formalism in Eq. (1), and by assuming noise-free kinematic coherent diffraction to be Gaussian distributed, compute the variance of standardised correlation coefficients obtained from a normalised cross-correlation of a Gaussian filter with the raw diffraction data,

$$V(\gamma,k,Q) = \left\{ \langle \gamma^2(I(\mathbf{r},x,y,Q),t)\rangle - \langle \gamma(I(\mathbf{r},x,y,Q),t)\rangle^2 \right\}_k. \tag{2}$$

where $\gamma$, the correlation coefficient is obtained from:

$$\gamma(x,y) = \frac{\sum_{x,y}\left(I(x,y)-\bar{I}_{u,v}\right)\left(t(x-u,y-v)-\bar{t}\right)}{\sqrt{\sum_{x,y}\left(I(x,y)-\bar{I}_{u,v}\right)^2\left(t(x-u,y-v)-\bar{t}\right)^2}}. \tag{3}$$

In Eq. (3), $t$ is the Gaussian filter, $\langle\ldots\rangle$ indicates averaging over $\mathbf{r}$, and $I$ is as in Eq. (1) with the exception that here the variance is computed on a pixel by pixel basis $(x, y)$ through the diffraction pattern stack resulting in a variance map which is thereafter azimuthally averaged (represented by $\{\ldots\}_k$) to obtain the variance as a function of scattering vector $k$. Eq. (2) is a standard expression for variance which can be found in any statistical reference manual and is easily recognised by the mnemonic "mean of the square minus square of the mean"; it differs in form from Eq. (1) by the change of variable and the normalisation factor in the denominator; we normalise our data through Eq. (2) prior to computing the variance. Whilst the Gaussian filtering was initially intended to mitigate noise, it became apparent that a change of variable in the variance calculations from intensity to a normalised score of the intensities structural significance simultaneously removed noise, background and standardised the datasets. In Eq. (3), $I(x,y)$ denotes the intensity value of the diffraction pattern at the point $(x,y)$, $\bar{I}_{u,v}$ is the mean value of $I(x,y)$ within the area of the Gaussian filter $t$ at the point $(x,y)$, and $\bar{t}$ is the mean value of the Gaussian filter. The denominator in Eq. (3) contains the variance of the zero mean diffraction pattern function $I(x,y)-\bar{I}_{x,y}$ and the zero mean Gaussian filter function $t-\bar{t}$ at the point $(x,y)$. The 2-D normalised cross correlation is used here as a standardised means to evaluate the significance of the raw diffracted intensity at each point in the diffraction pattern, and is scored upon the similarity of the local distribution of intensity around each point to our model diffraction maxima centred on that point. Our model diffraction maximum is a 7 × 7 pixel rotationally symmetric normalised Gaussian filter with a two pixel standard deviation, obtained by fitting a 2-D Gaussian function to a sharp Bragg spot in a diffraction pattern from the same sample series, which had crystallised after a 800 °C heat treatment. Using this approach, the absolute magnitude of the scattered intensity is irrelevant and instead, it is the shape of the intensity distribution around each pixel that becomes relevant, allowing coherent diffraction with poor SNR to emerge from the background. The resulting normalised correlation map is then a standardised transform of the diffraction pattern into a map of the diffracted intensity's structural significance, where scores range in value between −1 (maximally anti-correlated), zero (uncorrelated) and 1 (maximally correlated). Only positive scores are deemed structurally significant as we assume coherent diffraction to be approximately Gaussian, whereas the diffuse background and single pixel events are not Gaussian distributed. We thus remove the negatively scored intensity contributions from calculations which anti-correlate with our model Gaussian filter, and we assume that much of the noise and diffuse background in the system will result in this negative range of scored intensities. As a result of this normalisation, $\gamma(x,y)$ is invariant to brightness or contrast variations in the diffraction patterns (including from diffuse inelastic scattering), which are related to the values of the mean and the standard deviation; this has the effect of standardisation of the data-sets and preservation of real diffraction spots deemed structurally significant through positive correlation, whilst rejecting single pixel noise or X-ray events. Intuitively, this approach seems well suited to the study of structural change due to thermal annealing; each diffraction pattern across the sample series is rendered statistically equivalent by the transformation of raw intensity pixel values to a normalised correlation coefficient which itself is determined by the similarity of that pixel and its surrounding area (size of the filter) to the Gaussian filter centred on that pixel. As the Gaussian filter is a representation of diffraction maxima from the specimen that has attained long range order by thermal annealing at 800 °C, it serves as a reference in which the distribution of diffracted intensity in the remaining thermal annealing series, which have not attained that level of order, can be meaningfully compared using the relative values of the assigned correlation coefficients, which describe the degree structural significance that the raw intensity values contribute to the diffraction pattern. Inspired by methods used in image analysis for feature tracking, this approach renders the data invariant to inconsistencies in the illumination conditions during data acquisition [38], and removes the need to normalise after the variance is computed. It has also been shown that reliable extraction of information using the formal FEM technique is highly dependent upon the quality and reproducibility of the experimental data, and as such, steps have been laid out to accurately identify and correct artefacts in affected datasets [39] which we review in relation to our data.

Early work in scanned diffraction used a very small number of diffraction patterns and was very slow to collect [18] due to the limitations of the imaging detectors available at the time. More recently, rapid advances were made in high speed acquisition of scanned diffraction data using an optically coupled camera [40–42]. Whilst these advances were principally made with precession electron diffraction in mind [41], spatial resolution can be increased by deactivating the precession and this is then ideal for application in FEM with a resolution of ~1 nm, although the noise in the system requires some special treatment, as described below. Future work will be further aided by the use of a new generation of direct electron detectors that allow high frame rate collection (~1 kHz) with almost perfect quantum efficiency [43–45].

In addition to the classic variance-based analysis of the scanned diffraction data [18], it is also possible to image features of the scanned diffraction dataset in real space due to the spatially indexed nature of the diffraction pattern collection which we discuss further in Section 3.2. Using the Virtual Dark-Field Imaging (VDFi) approach of Rauch and Véron, real space images are formed from the indexed scanned diffraction dataset by using the intensity in selected parts of each diffraction pattern [46]. This is a counterpart to classic dark field imaging, where images are formed using only selected scattered electrons instead of the undiffracted central spot, and this is realised by the insertion of a physical aperture in the back focal plane of the objective lens. This has the effect of masking almost all the diffraction pattern, except the one reflection which is visible through the aperture. For the VDFi construction [47], the whole back focal plane is collected with a spatially resolved detector and later numerically reconstructed to rebuild real space representations of the materials, using any desired combination of integrated intensity (unlimited combinations of numerical apertures can be employed) out of the diffraction patterns, in order to create the pixel values of the reconstructed real-space images. This has one clear advantage over conventional dark field imaging: the whole back focal plane is collected allowing any number of different dark field images to be reconstructed afterwards, whereas conventional dark field imaging has just one reflection selected per image. Of course, this comes with a downside that a vast quantity of data needs to be collected to map any reasonable scan area at a sensible spatial resolution, when only a tiny fraction of the data is used in the production of each VDFi image. Thus far the technique has been used to spatially map phases and orientations of polycrystalline materials [46, 48]; here we extend the concept of FEM to map the evolution of medium range order and nano-crystallite nucleation in glasses as a function of thermal annealing, effectively spatially resolving the FEM results. Instead of a two dimensional plot of variance as a function of scattered intensity $I(k)$, we can now visualise the entire dataset simultaneously as a spatially resolved variance map at a given scattering vector (or indeed using any part of the scattered beam). This is similar to FEM using dark-field imaging, in which the intensity is collected in the form of real space images, which have one scattering vector $k$, one probe size $Q$, and many spatial samples $r$, $k$ and $Q$ get changed, and the next image is acquired. On the contrary, in our method of STEM nano-diffraction, $I(\mathbf{r}, k, Q)$ is obtained in the form of an electron diffraction pattern acquired with a nanometre-sized probe, which has one $Q$, one $\mathbf{r}$, and many $k$ values. That probe is then rastered across the sample to acquire many $\mathbf{r}$ [49]. VDFi therefore enables FEM data acquired via scanned diffraction to be quantified in new ways, by combining the real space and reciprocal space representations in one analysis. This combined approach is used in the present work to probe the changes in amorphous $Ta_2O_5$ films on thermal annealing.

## 2. Experimental

### 2.1. Samples/annealing process/specimen preparation

The amorphous IBS tantala coatings were manufactured by the Commonwealth Scientific and Industrial Research Organisation (CSIRO, Materials Science and Engineering Division, West Lindfield, NSW, Australia). The coatings were deposited onto fused silica substrates and were subject to post-deposition annealing at 300 °C, 400 °C and 600 °C for 24 h in air. Specimens were prepared for the scanning electron diffraction studies using a standard cross section method of gluing two sections of film face to face, encapsulation in a brass tube with epoxy resin, slicing, polishing and dimpling. The samples were then thinned to electron transparency using a Gatan Precision Ion Polishing System (PIPS) (Gatan Inc., Pleasanton, CA, (USA), which used Ar + ion irradiation at a relatively low energy, 4 kV beam, and a final 0.5 kV polishing stage, to avoid any changes to the sample material structure.

### 2.2. Data collection and reduction

The electron diffraction data were collected on a JEOL JEM2100F operating at 200 keV at the University of Grenoble. The microscope was set to nano-beam mode in TEM, encapsulating diffraction from a volume set by the beam diameter, 2 nm FWHM in this work, with a dedicated hardware solution made by NanoMEGAS which controlled the deflection coils of the TEM allowing it to be operated in a ~2 mrad low convergence angle STEM mode [40, 42, 50]. Although we did not measure the probe current absolutely, it was kept low enough that there was no speckle movement under the stationary beam, which indicates that there is no modification of the atomic structure of the sample by the beam. The diffraction patterns were recorded using an external video camera imaging the TEM phosphorous screen with a dynamic range of 8 bits, at a rate of 100 patterns per second over the area of interest. All diffraction patterns were linear unsaturated measurements of the diffracted intensity.

At this probe resolution of ~2 nm, the resulting diffraction patterns from amorphous tantala take the form of speckly distributions of diffracted intensity as depicted in Fig. 1(a), as is typical of such FEM experiments, as opposed to the diffuse rings that are conventionally associated with diffraction from larger volumes of amorphous materials seen in Fig. 1(b) (recorded with an exposure time an order of magnitude greater than that of Fig. 1(a), from a volume five orders of magnitude larger). Of course, simply adding up many speckle patterns from different regions would produce a pattern like Fig. 1(b). Whilst the authors are aware of methods that explicitly investigate the length scale of MRO using a range of probe sizes [12], these were not explicitly used in this work, and the 2 nm probe size was found to optimal for FEM studies on this sample with this microscope.

The optical coupling in this work, whilst advantageous for rapid data acquisition in low convergence angle STEM mode, has the undesired effect of introducing artefacts and noise into the data. This includes artefacts such as spurious X-rays impinging on the phosphor screen, as well as significant electronic noise in the camera. Methods investigated to reduce these, as well as other sources of noise led to the development of the version of FEM used in this work. Fig. 2 shows an example of a typical diffraction pattern used here for analysis, before and after the data reduction process.

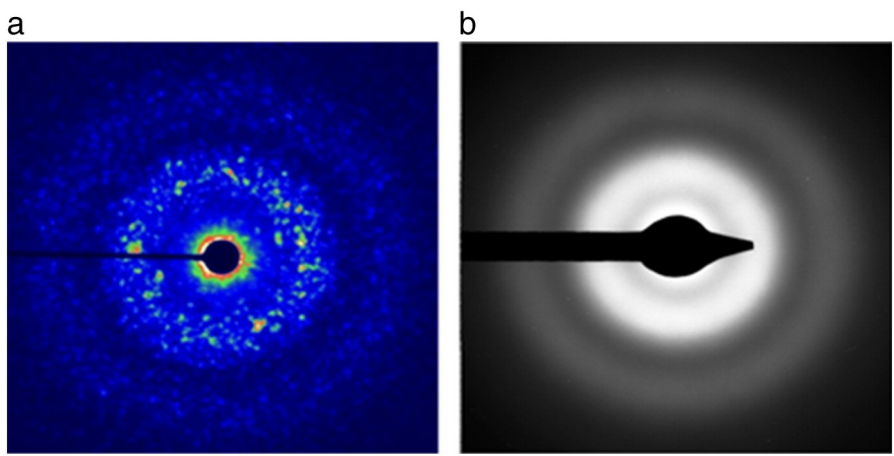

**Fig. 1.** (a) Typical tantala FEM diffraction pattern using 2 nm probe FWHM, (b) typical tantala diffraction pattern using 50 nm probe FWHM, recorded on a CCD.

Although the general purpose of the 2D normalised cross correlation is for the determination of the position of a feature within an image using template matching, all of the features of interest in the diffraction pattern bear similarity to our chosen template. Thus the result is not a correlation map of the likeliest position of a feature, but a correlation map of the similarity of regions of the diffraction pattern to the Gaussian filter. Although the variance Eqs. (1) and (2) are not mathematically equivalent, they both provide a dimensionless value for the variance which are directly related to the structural heterogeneity of the samples. Therefore, we compute the variance of the correlation coefficients of the diffracted intensity with a 2-D Gaussian as opposed to the directly detected intensity values, and use them as a proxy for the scattered intensity by approximating the two dimensional distribution of intensity to be Gaussian in nature for individual diffraction maxima. The use of a constant Gaussian filter throughout the whole data series is key in obtaining meaningful comparison of the standardised data sets of the Ta$_2$O$_5$ heat treatment series, and serves as the model which is used to select structurally significant regions of diffracted intensity for calculations.

Li et al. [39] provided a basis upon which FEM can be performed with confidence in the quantitative magnitude of the data, which entails the identification and removal of common artefacts in the dataset. We examine these noise sources here in relation to our data.

The datasets used in this work were from conventionally prepared cross sectional samples which contained some thickness variations, although care was taken to choose areas for analysis which were extremely thin and with very small thickness variations. The consequence was that no saw-tooth pattern was visible in a plot of $I(k)$ versus pattern number at high $k$ before or after data reduction (which would indicate a large thickness change). No voids were identified, neither were any roughness effects. A small amount of carbon contamination was visible in the raw data in a plot of $I(k)$ versus pattern number at high $k$, although after data reduction the effect vanished. Furthermore, no multiple scattering effects or large nanocrystals were observed.

Additionally, we observed no noticeable dependence of the variance upon the specimen thickness, which was determined by plotting the main variance peaks of 56 FEM signals (100 diffraction patterns each), from parallel lines normal to the small thickness gradient in the dataset, for both the 300 °C and 600 °C data sets. These results confirm that any differences in the volumes of the samples studied were inconsequential to our analysis [39]. The 400 °C data was sampled from areas similar to the 300 °C and 600 °C data sets and underwent identical sample preparation, indicating minimal thickness variation within each sample and from sample to sample. The usual FEM signal is also somewhat sensitive to probe coherence; however using a small semi-convergence angle, and by performing calculations only where scattered intensities have been determined to be structurally significant, as opposed to the total scattered intensity, this sensitivity is reduced.

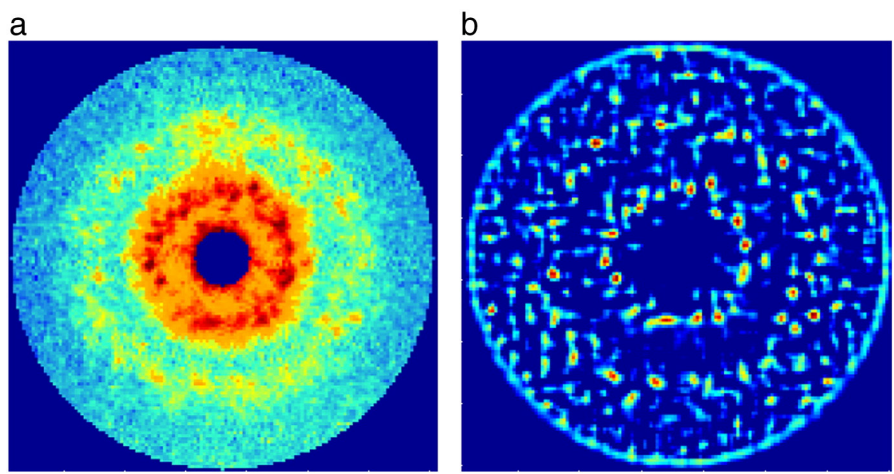

**Fig. 2.** (a) Raw diffraction pattern, (b) diffraction pattern cross correlated with normalised Gaussian filter, retaining only positive correlations. The central beam has been removed for clearer comparison of features.

## 3. Analysis and results

### 3.1. Fluctuation Electron Microscopy (FEM)

Fig. 3 shows the results of the average variance of five data sets, each using 1000 diffraction patterns, computed for the 300 °C, 400 °C and 600 °C samples, with the error bars showing the standard deviation of the variance computed over the five data sets. These datasets simultaneously confirm the presence of MRO and of a change in short range atomic structure, both as a consequence of thermal annealing.

As mentioned previously, the atomic structure of these coatings annealed up to 600 °C appears homogeneous when probed with a beam of ≥50 nm diameter as a function of scattering vector and position on the sample; so unlike the peaks in Fig. 3, a variance plot at that resolution would appear relatively flat and give no hint to any underlying medium range order. Here, probing the structures with a 2 nm diameter probe over thousands of areas of the samples, and obtaining these variance peaks, confirms the heterogeneity of the atomic structure over these length scales. There is always a trade-off between spatial resolution and averaging when selecting a probe size and this technique is effectual only when the probe is sufficiently small that averaging of the volume does not remove resolution in reciprocal space of the diffracted intensity. The 2 nm probe here was selected qualitatively based upon the prominence and contrast of the observed variation in diffracted intensity at different positions on the sample. And whilst a different level of heterogeneity can be observed using different probe sizes, it has been shown that there exists a peak in the heterogeneity of the structure that coincides with the maximum variance, and is observed when the probe size is of comparable dimension to the characteristic length scale of the order within the material [12]. Gibson et al. developed a technique which can quantitatively determine this characteristic length scale using different probe sizes [12] which we plan to use in future studies.

At this scale, the change in atomic structure due to thermal annealing is apparent; the magnitude of the 600 °C variance is significantly larger than that of the 300 °C and 400 °C data, indicating further deviation from homogeneity that can be explained in terms of the structure beginning to organise itself into regions of greater order. This seems to correlate with mechanical loss measurements of the same materials which show no significant difference between the 300 °C and 400 °C coatings, but shows significant changes between those and the 600 °C coating [7].

### 3.2. Virtual Dark-Field Imaging

The concept behind the Virtual Dark-Field Imaging (VDFi) technique is shown in Fig. 4. Each of the circles represents a diffraction pattern whose resolution $Q$ is dictated by the size of the incident electron probe (2 nm FWHM), rastered across the sample surface. Each of the 1.5 nm × 1.5 nm squares represent a pixel of the reconstructed VDFi image and are the step size in which the electron beam is rastered. The intensity of each pixel is the integrated intensity of any part of the diffraction pattern selected. Fig. 4(a) shows the selection of a line segment containing three Bragg reflections; integrating the resultant intensity from the same location in each diffraction pattern, we obtain an image highlighting the persistence of that structural motif in such a volume over a user defined area. Likewise shown in Fig. 4(b), (c) and (d), any numerical aperture or combination thereof, can be created to enable investigation of the persistence of such nanoscale order over any desired area.

Fig. 5 shows an example of VDFi images of the 300 °C, 400 °C and 600 °C samples created using a virtual annular aperture such as in Fig. 4(c), integrated over the scattering vectors between 0.287 Å⁻¹ and 0.300 Å⁻¹. The effect of this range selection is to visualise the FEM data in real space where the amorphous structure is furthest from homogeneity, which in this case predominantly describes the heterogeneity of the tantalum–tantalum nearest neighbour environment. Setting the angular range for VDFi at some other point where the variance is less will show less in the way of interesting structure. It is important to note that these VDFi images were constructed using the same data sets that were used in the FEM analysis, and are thus consistent with the FEM results and investigate the spatial persistence of fragments responsible for medium range ordering.

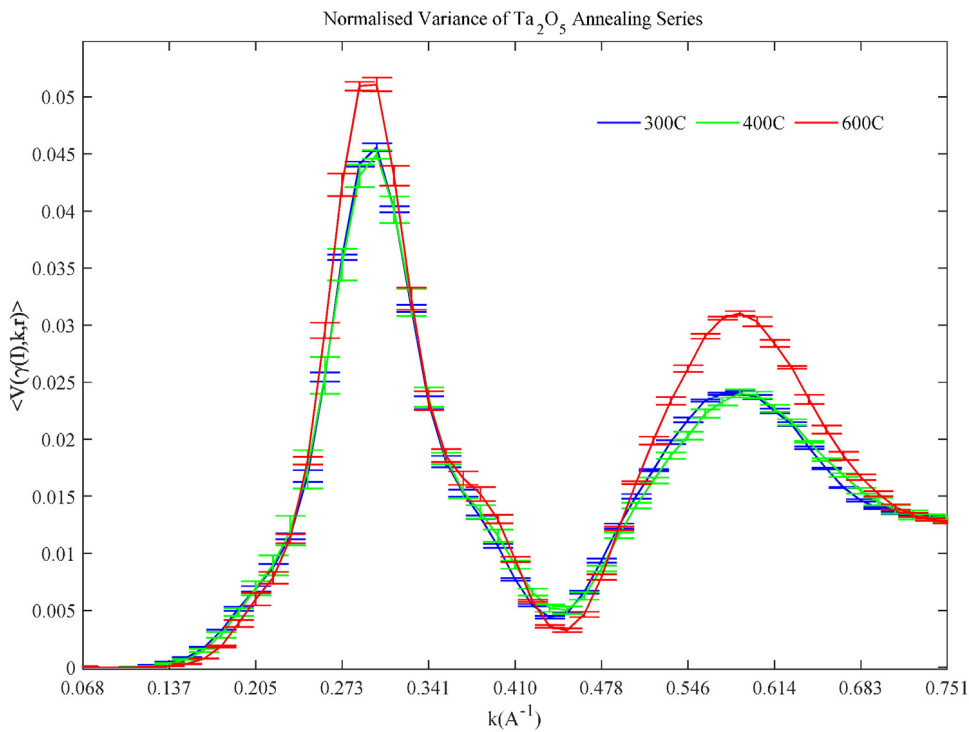

**Fig. 3.** Comparison of the normalised variance of the tantala thermal annealing series. The variance plot shows that a similar magnitude of change between the first and second peaks is seen as a function of heat-treatment.

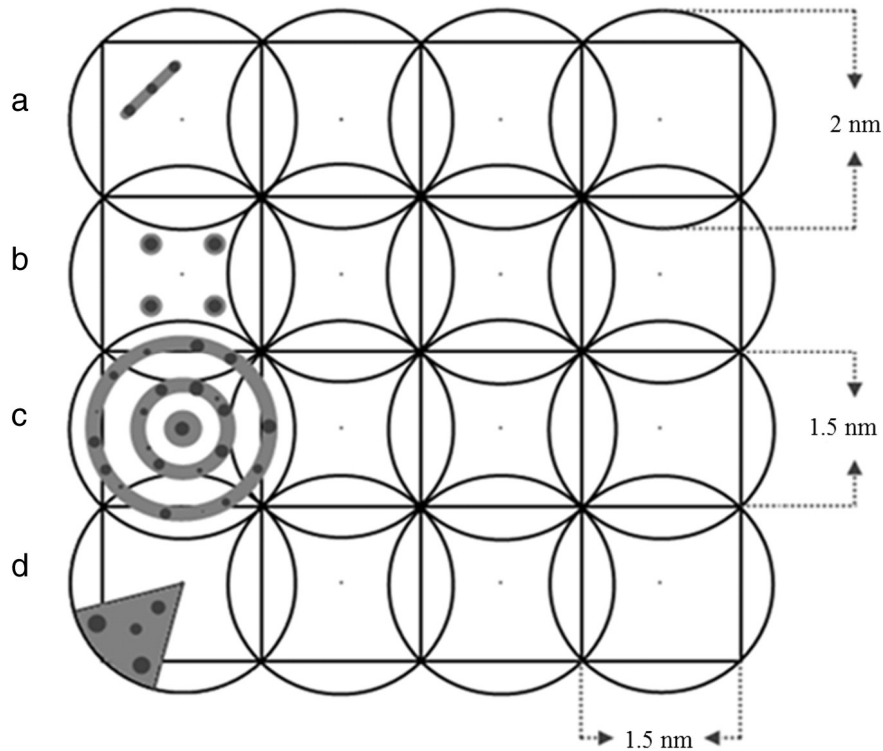

**Fig. 4.** Schematic of the Virtual Dark-Field Imaging process. Circles represent a diffraction pattern from a nano-volume, squares represent pixels in the VDFi image. Examples of virtual apertures (a) line segment, (b) beam spots, (c) annuli, (d) sector.

As intimated before, these virtual images can be constructed from the integrated intensity of annuli at all scattering vectors and allow the different changes in structure due to thermal annealing to be charted as a function of such. The results of the Gaussian functions fits to the histograms of intensity distributions of the virtual images shown in Fig. 5 complement that of the FEM study. As heat treatment temperature increases, the mean intensity of the main diffraction peaks increases, with the relative incidence of higher diffracting centres decreasing as would be expected with certain areas becoming more ordered. In this instance we still retain the spatial variance of the diffracted intensity at that scattering vector and in each VDFi image the groupings of red pixels indicate areas of the greatest diffracted intensity, corresponding to the highest degree of structural ordering. In all of these VDFi images, clusters of red pixels are found. Whilst many of the clusters are just 2 pixels wide (about 3 nm), some are as large as 4 pixels across, suggesting patches of order to greater than 5 nm in places.

The results of Fig. 5(a, b, c) (where pixel colours represent the range of integrated intensity and have a one to one correspondence in each of the VDFi images) are quantified in Fig. 5(d), which is a plot of the frequency of occurrence versus mean intensity. Whilst Fig. 5(a) and (b) for the 300 °C and 400 °C annealed materials show a rather similar frequency distribution, as would be expected from their very similar variance plots in Figs. 3, 5(c) shows that the peak of the distribution is moved to higher intensities, although with a lower height. This is fully in accord with the changes in variance observed at the same scattering vector in Fig. 3, and demonstrates an increased ordering after annealing at 600 °C into fewer, larger ordered regions.

## 4. Discussion

We note that the mechanical loss of these coatings at low temperatures increase with increasing heat treatment [7], and thus it seems plausible that the reason may be related to degree of medium range order. Moreover, the shift of the distributions in Fig. 5(d), due to thermal annealing, very closely resembles the shift of the mechanical loss peaks measured for these same samples [7] and will be investigated further to ascertain if any correlations of the mechanical loss with the presence of nano-crystallites are found. Recent work by Treacy and Borisenko on the local structure of amorphous silicon found para-crystalline structures containing local cubic ordering at the 10 to 20 Å length scale [12]. Similarly, our results suggest that the structure of amorphous tantala can be better described by a phase separated heterogeneous model than the continuously uniform random network model of Zachariasen [51]. Other studies have shown favour to this same nano-crystalline theory of glass [52–54]. Although similar results to the study in [54] initially have been interpreted in terms of Zachariasen's continuous random network model [55], further studies revealed the presence of nanoscale ordering [56].

It is important to consider the connection between the structural trends highlighted as a consequence of annealing in this work, and the trends in mechanical loss with annealing for the same materials. It is found that increased annealing temperature to 600 °C reduces the room-temperature mechanical loss, making the use of 600 °C annealing advantageous for mirrors designed for use at room temperature [7] (such as Advanced LIGO). At the same time, this 600 °C annealing causes the appearance of a significant and well-localised peak in loss at around 20 K, making this annealing possibly less suited for the preparation of coatings for use in cryogenic detectors [7], such as Kagra, currently under construction in Japan [57], and the proposed Einstein Telescope [9]. It may be suggested that the increased ordering affects the number and energetics of defects forming double well potentials at boundaries between paracrystallites. These defects appear to have relatively low energy barrier between two metastable states, reducing the overall losses at elevated temperatures whilst increasing the losses at lower temperatures, similarly to the mechanical loss mechanism proposed for amorphous silica [58].

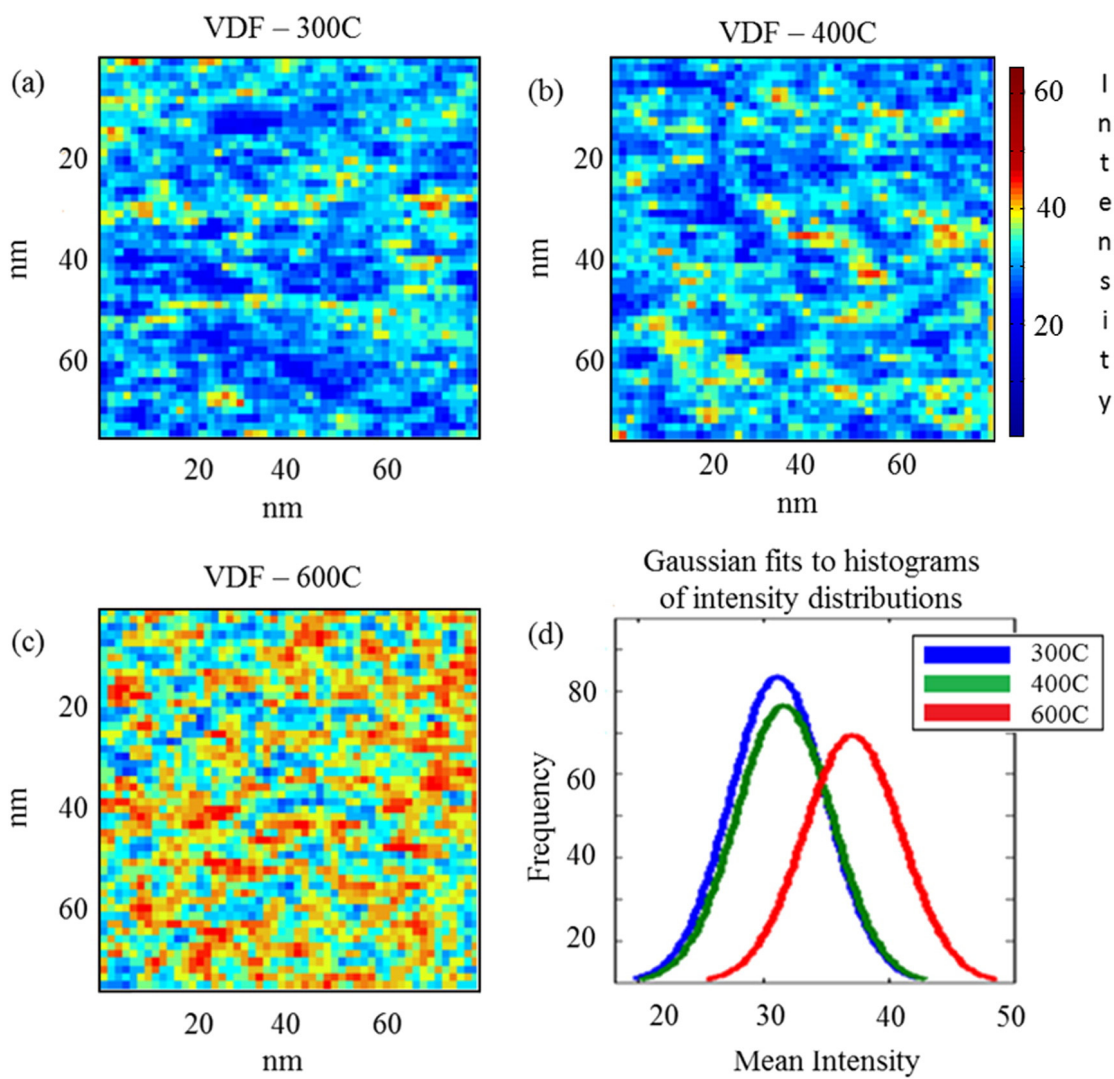

**Fig. 5.** Virtual Dark-Field Images of (a) 300 °C, (b) 400 °C, (c) 600 °C from the tantala thermal annealing series. (d) Gaussian fits to the histograms of intensity distributions of the VDFi images. The pixel intensities represent integrated intensities of annuli in the range 0.287 Å$^{-1}$ to 0.300 Å$^{-1}$. (For interpretation of the references to colour in this figure, the reader is referred to the web version of this article.)

Future work will further explore the proposed VDFi approach in conjunction with FEM to look for the evolution of structural motifs that correlate with the measured mechanical loss of these coatings resulting from alloying and thermal annealing. It is also planned to perform improved versions of this experiment using the JEOL ARM200F at the University of Glasgow, imaging onto a Medipix III, electron-counting direct electron detector [43–45], in order to study the medium range order in titania-doped tantala.

## 5. Summary

We have shown using FEM and VDFi that medium range order exists in Ion Beam Sputtered amorphous tantala thin film mirror coatings on the length scale of 1.5–6 nm (from min/max cluster sizes in VDFi data), and that said order increases with increasing levels of thermal annealing, showing definitively for the first time that changes in the medium range ordering of the atomic structure of amorphous tantala result from thermal annealing. It is moreover demonstrated that Virtual Dark-Field Imaging is a valuable spatially resolvable FEM technique for the study of the evolution of local ordering in glasses due to thermal annealing. It seems that the increased medium range order with annealing to 600 °C is correlated with both the reduction in room temperature mechanical loss and the increase in low temperature loss around 20 K, and a tentative model for the mechanism underlying this correlation has been proposed.

## Acknowledgements

MJH, IM, IWM and SR gratefully acknowledge the support of the STFC (*ST*/L000946/1 'Investigations in Gravitational Research'), for this work. MJH is grateful to the EPSRC for a PhD studentship. IWM is supported by a Royal Society Research Fellowship. The authors would like to thank Dr. Stavros Nicolopoulos of NanoMegas SPRL for his support for the work in Grenoble, without which this work would not have been possible. RB and MMF gratefully acknowledge the support of the NSF, under award number PHY-1404430.